\documentclass[10pt,twocolumn,aps,prd,longbibliography,nobibnotes,nofootinbib,floatfix]{revtex4-2}
\usepackage{amsmath,amssymb}
\usepackage{booktabs}
\usepackage{hyperref}

\hypersetup{
  colorlinks=true,
  linkcolor=blue,
  citecolor=magenta,
  pdftitle={Black Hole Entropy Beyond the Wald Term in Nonminimally Coupled Gravity: A Covariant Phase Space Decomposition},
  pdfauthor={}
}

\allowdisplaybreaks[2]
\newcommand{\Lie}{\mathcal{L}}
\newcommand{\dd}{\mathrm{d}}
\newcommand{\SW}{S_{\mathrm W}}
\newcommand{\SQ}{S_{\mathrm Q}}
\newcommand{\SH}{S_{\mathrm H}}
\newcommand{\Sone}{S_1}
\newcommand{\DeltaS}{\Delta S}
\newcommand{\RhX}{\widehat{\mathcal R}_X}
\newcommand{\RhW}{\widehat{\mathcal R}_W}

\newcommand{\Lagr}{\mathbf L}
\newcommand{\Thetaform}{\mathbf\Theta}
\newcommand{\ThetaS}{\mathbf\Theta_{\rm S}}
\newcommand{\omegaform}{\boldsymbol\omega}
\newcommand{\Jxi}{\mathbf J_\xi}
\newcommand{\Qxi}{\mathbf Q_\xi}
\newcommand{\QS}{\mathbf Q^{\rm S}_\xi}

\newcommand{\QdS}{\mathbf Q^{\rm S}_{\delta\xi}}
\newcommand{\Bhor}{\mathcal H}
\newcommand{\kxi}{\mathbf k_\xi}
\newcommand{\kS}{\mathbf k^{\rm S}_\xi}
\newcommand{\Wm}{\mathcal W_{\rm m}}
\newcommand{\Wpar}{\mathcal W_{\rm par}}
\begin{document}
\title{Black Hole Entropy Beyond the Wald Term in Nonminimally Coupled Gravity: A Covariant Phase Space Decomposition}

\author{Jia-Zhou Liu$^{1,2}$}
\author{Shan-Ping Wu$^{1,2}$}
\author{Shao-Wen Wei$^{1,2}$}
\author{Yu-Xiao Liu$^{1,2}$}
\email{liuyx@lzu.edu.cn}

\affiliation{$^{1}$Lanzhou Center for Theoretical Physics, Key Laboratory of Theoretical Physics of Gansu Province,\\ Key Laboratory of Quantum Theory and Applications of MoE, \\Gansu Provincial Research Center for Basic Disciplines of Quantum Physics, \\ Lanzhou University, Lanzhou 730000, China}

\affiliation{$^{2}$Institute of Theoretical Physics $\&$ Research Center of Gravitation, \\School of Physical Science and Technology, Lanzhou University, Lanzhou 730000, China}

\begin{abstract}
We study the entropy of static, spherically symmetric black holes in
diffeomorphism-invariant theories with nonminimal matter--curvature couplings,
using the covariant phase space formalism. For regular bifurcate Killing
horizons, the Iyer--Wald construction gives the standard Wald entropy. If a matter field cannot be smoothly extended to the regular bifurcation surface, however, the entropy-sector horizon surface charge variation can contain finite contributions that are not included in the Wald entropy density. In the representative obtained by directly varying the action, and after ordinary non-gravitational boundary terms have been separated into the work sector, we decompose the entropy entering the first law of black hole thermodynamics as
\(\SH=\SW+\Sone+\DeltaS\). Here \(\SW\) is the Wald entropy, \(\Sone\) is the non-Wald part of the entropy-sector Noether charge, and \(\DeltaS\) is the remaining integrable part of the entropy-sector horizon surface charge variation. Applying this criterion to Kalb--Ramond, bumblebee, and extended Gauss--Bonnet black holes, we find that the regular Kalb--Ramond branch
has \(\SH=\SW\), the bumblebee branches yield either \(\Sone=0\) with
\(\DeltaS\neq0\) or a cancellation between \(\Sone\) and \(\DeltaS\), and the Weyl-vector extended Gauss--Bonnet examples require both corrections. This provides a direct test of whether the Wald density is sufficient or whether the full  horizon surface charge variation is required.
\end{abstract}

\maketitle

\section{Introduction}

In Einstein gravity, black hole thermodynamics and Hawking radiation fix the entropy to be \(S=A/(4G)\) \cite{Bekenstein:1973ur,Bardeen:1973gs,Hawking:1975vcx}. In theories with higher-curvature terms or nonminimal matter--curvature couplings, the area is no longer expected to be the universal entropy functional \cite{Jacobson:1993xs,Iyer:1994ys,Jacobson:1995uq}.

The standard framework for this question is the covariant phase space formalism, in which conserved charges arise as surface terms associated with diffeomorphism symmetries \cite{Crnkovic:1986ex,Zuckerman:1986vzu,Lee:1990nz,Compere:2006my}. The Iyer--Wald formalism gives the familiar black hole entropy and thermodynamic first law for regular bifurcate Killing horizons \cite{Wald:1993nt,Iyer:1994ys,Iyer:1995kg,Jacobson:1993vj}. For a diffeomorphism-invariant theory with Lagrangian density \(L\), the Wald entropy is the horizon integral of a local density,
\begin{equation}
	\label{eq:wald-standard}
	\SW
	=
	-2\pi\oint_{\Bhor} E_R^{abcd}\,\epsilon_{ab}\epsilon_{cd},
	\qquad
	E_R^{abcd}\equiv \frac{\partial L}{\partial R_{abcd}},
\end{equation}
where \(\epsilon_{ab}\) is the antisymmetric binormal to the horizon cross section \(\Bhor\), normalized by \(\epsilon_{ab}\epsilon^{ab}=-2\). In stretched horizon calculations, \(\Bhor\) denotes the limiting cross section. In Einstein gravity, Eq.~\eqref{eq:wald-standard} reduces to the area law, and in many higher-curvature theories it gives the entropy entering the thermodynamic first law \cite{Jacobson:1993xs,Iyer:1995kg,Jacobson:1995uq}.

The standard bifurcation-surface reduction is not always sufficient to determine the entropy-sector part of the horizon surface charge variation. That reduction uses a regular bifurcation surface, a fixed normalization of the Killing generator, and the standard horizon charge relation. In particular, the matter fields that enter the nonminimal curvature couplings must be smoothly extendable to the regular bifurcation surface. If a scalar, vector, or tensor field fails to do so, the usual \(\xi^a=0\) argument no longer guarantees that the terms proportional to \(\xi^a\), in particular the \(\mathbf W_a\xi^a\) part of the entropy-sector Noether charge and the \(\xi\cdot\ThetaS\) part of the surface charge variation, vanish in the horizon limit. This is the situation in the vector examples below: some one-form components are singular in coordinates that are regular at the bifurcation surface. Einstein--Horndeski black holes provide explicit examples in which the entropy entering the first law contains a contribution beyond the standard Wald entropy \cite{Feng:2015oea,Feng:2015wvb,Minamitsuji:2023nvh}. Similar subtleties in vector-tensor black hole thermodynamics have also been noted in nonminimally coupled vector and generalized Proca models \cite{Fan:2017bka,Minamitsuji:2024ygi,An:2024fzf}.

This issue should be distinguished from the usual Jacobson--Kang--Myers (JKM) ambiguity of the covariant phase space representative. Related JKM-invariant entropy constructions address dynamical and second-law questions for higher-curvature theories with nonminimally coupled fields \cite{Jacobson:1993vj,Wall:2024lbd}. The stationary first-law problem addressed here is different. We fix the representative defined by the displayed Lagrangian and by the presymplectic potential obtained from its direct variation. The aim is to decompose the entropy entering the first law into the part already contained in the Wald density, the non-Wald part of the entropy-sector horizon Noether charge, and the remaining integrable contribution from the entropy-sector presymplectic potential.

In this fixed representative, and after subtracting the ordinary non-gravitational and parameter-work terms, the entropy entering the first law decomposes as
\begin{equation}
	\label{eq:entropy-main-result-intro}
	\SH=\SW+\Sone+\DeltaS.
\end{equation}
Here \(\SW\) is the standard Wald entropy, \(\Sone\) is the correction from the non-Wald part of the entropy-sector Noether charge, and \(\DeltaS\) is the integrable remainder of the entropy-sector horizon surface charge variation. These terms are defined precisely in Secs.~\ref{sec:cps-setup} and~\ref{sec:horizon-decomposition}.

A useful way to organize the horizon calculation is to separate the presymplectic potential part of the horizon variation. We restrict attention to static, spherically symmetric black hole families, for which the horizon generator is taken to be \(\xi^a=(\partial_t)^a\). No rotational Killing field \((\partial_\phi)^a\) or angular momentum term is included. For theories of the form
\begin{equation}
	\label{eq:intro-lagrangian-class}
	\Lagr=\boldsymbol{\epsilon}\,L(g_{ab},R_{abcd},\phi^I,\nabla\phi^I),
	\qquad
	\delta\xi^a=0,
\end{equation}
and denoting the horizon surface gravity by \(\kappa\), the presymplectic potential contribution at the horizon can be written as
\begin{equation}
	\label{eq:intro-theta-decomposition}
	-\oint_{\Bhor} \xi\cdot\ThetaS
	=
	-\frac{\SW+\Sone}{2\pi}\,\delta\kappa+\mathcal R_X+\mathcal R_W.
\end{equation}
Here \(\SW+\Sone=\SQ\) is the entropy read from the entropy-sector horizon integral of the Noether charge.
The remainder \(\mathcal R_X\) comes from horizon components of \(E_R^{abcd}\nabla_d\delta g_{bc}\) not contained in the Wald entropy density, while \(\mathcal R_W\) is the remaining entropy-sector presymplectic potential contribution associated with the complete \(\mathbf W_a\xi^a\) charge contribution after its \(-\Sone\delta\kappa/(2\pi)\) part has been separated. In explicit models \(\mathcal R_W\) is evaluated from \(\ThetaS\), rather than from a single local curvature-derivative term.

We then apply the decomposition to three classes of black hole solutions.
The Kalb--Ramond example is nonminimal but still has \(\SH=\SW\): the nonminimal coupling changes the Wald entropy itself, but the remainders vanish.
Bumblebee gravity gives two complementary vector-tensor branches: a spacelike branch with \(\SW=\SQ\) but \(\DeltaS\neq0\), and a lightlike branch with \(\Sone\neq0\) and \(\DeltaS\neq0\) that cancel in \(\SH\).
Extended Gauss--Bonnet gravity in Weyl geometry provides a higher-curvature example with a Weyl vector, where the curvature dependence of \(E_R^{abcd}\) and the vector both enter the horizon calculation.

The paper is organized as follows.
Section~\ref{sec:cps-setup} reviews the covariant phase space identities and fixes the horizon charge notation.
Section~\ref{sec:horizon-decomposition} derives the stationary entropy decomposition and the horizon decomposition of the presymplectic potential contribution.
Section~\ref{sec:models} applies the formalism to Kalb--Ramond, bumblebee, and extended Gauss--Bonnet black holes.
Section~\ref{sec:conclusion} summarizes the entropy decomposition and the resulting criterion for when the standard Wald expression is sufficient.

\section{General covariant phase space setup}
\label{sec:cps-setup}

\subsection{Basic covariant phase space identities}

Consider a \(D\)-dimensional diffeomorphism-invariant theory with Lagrangian \(D\)-form
\begin{equation}
	\label{eq:lagrangian-form}
	\Lagr
	=
	\boldsymbol{\epsilon}\,
	L\bigl(
	g_{ab},R_{abcd},\phi^I,\nabla\phi^I
	\bigr),
\end{equation}
where \(\phi^I\) stands collectively for the scalar, vector, and tensor fields in the theory, and \(\Phi\) denotes the full set of dynamical fields.
Its variation has the form
\begin{equation}
	\label{eq:variation-form}
	\delta \Lagr
	=
	\mathbf E[\Phi]\delta\Phi
	+ \dd \Thetaform[\Phi,\delta\Phi].
\end{equation}
In the covariant phase space formalism, \(\delta\) is the exterior variation on field space, \(\mathbf E[\Phi]=0\) are the Euler--Lagrange equations, and \(\Thetaform[\Phi,\delta\Phi]\) is the presymplectic potential \((D-1)\)-form. Its antisymmetrized second variation defines the presymplectic current
\begin{equation}
	\label{eq:symplectic-current}
	\omegaform[\Phi;\delta_1\Phi,\delta_2\Phi]
	=
	\delta_1\Thetaform[\Phi,\delta_2\Phi]
	- \delta_2\Thetaform[\Phi,\delta_1\Phi].
\end{equation}
Its integral over a hypersurface \(\Sigma\) gives the presymplectic form \(\Omega_\Sigma\). When the corresponding surface charge variation is integrable, it defines the Hamiltonian generator associated with the symmetry~\cite{Lee:1990nz,Iyer:1994ys,Compere:2019qed}.

Because \(\Lagr\) is a diffeomorphism-invariant top-form, a variation generated by \(\delta_\xi\Phi=\Lie_\xi\Phi\) satisfies
\begin{equation}
	\label{eq:diffeo-variation}
	\delta_\xi \Lagr
	=
	\Lie_\xi \Lagr
	=
	\dd(\xi\cdot\Lagr).
\end{equation}
Comparing Eqs.~\eqref{eq:diffeo-variation} and \eqref{eq:variation-form} gives the Noether identity
\begin{equation}
	\label{eq:noether-identity}
	\mathbf E[\Phi]\Lie_\xi\Phi + \dd\Jxi = 0.
\end{equation}

For a vector field \(\xi^a\), the Noether current is
\begin{equation}
	\label{eq:noether-current}
	\Jxi
	=
	\Thetaform[\Phi,\Lie_\xi\Phi] - \xi\cdot\Lagr,
\end{equation}
so on shell it is closed. By Noether's second theorem and the local Poincar\'e lemma, it is therefore locally exact~\cite{Avery:2015rga,Compere:2019qed,Barnich:2018gdh,Barnich:2000zw}:
\begin{equation}
	\label{eq:noether-charge}
	\Jxi = \dd \Qxi.
\end{equation}
Here \(\Qxi\) denotes the local Noether charge \((D-2)\)-form associated with \(\xi^a\).
Varying Eq.~\eqref{eq:noether-current} then yields the standard surface charge identity
\begin{equation}
	\label{eq:surface charge-identity}
	\omegaform[\Phi;\delta\Phi,\Lie_\xi\Phi]
	=
	\dd\kxi[\Phi;\delta\Phi,\Lie_\xi\Phi],
\end{equation}
where
\begin{equation}
	\label{eq:kxi-def}
	\kxi
	=
	\delta\Qxi-\xi\cdot\Thetaform[\Phi,\delta\Phi].
\end{equation}
This form assumes a field-independent generator, \(\delta\xi^a=0\) \cite{Lee:1990nz,Iyer:1994ys,Compere:2019qed}. Let \(\Sigma\) have outer boundary \(S_r\) and inner boundary \(\Bhor\).

As usual, \(\Thetaform\) is defined only up to the addition of an exact form. The resulting ambiguity structure in \(\omegaform\), \(\Qxi\), and the entropy functional is the standard JKM ambiguity \cite{Iyer:1994ys,Jacobson:1993vj}. In the covariant phase space construction one may shift
\begin{equation}
	\label{eq:jkm-lagrangian-potential-shift}
	\begin{aligned}
		\Lagr\rightarrow \Lagr+\dd\boldsymbol{\mu},
		\qquad
		\Thetaform\rightarrow
		\Thetaform+\delta\boldsymbol{\mu}+\dd\mathbf Y,
	\end{aligned}
\end{equation}
which induces corresponding shifts of the Noether charge,
\begin{equation}
	\label{eq:jkm-charge-shift}
	\Qxi\rightarrow
	\Qxi+\xi\cdot\boldsymbol{\mu}
	+\mathbf Y[\Phi,\Lie_\xi\Phi]+\dd\mathbf Z_\xi .
\end{equation}
These transformations change the local representatives of \(\Thetaform\), \(\omegaform\), and \(\Qxi\), rather than introducing independent physical terms in the first law. Throughout this paper we fix the JKM representative by taking the Lagrangian \(D\)-form as written and the presymplectic potential obtained from its direct variation; no additional \(\dd\boldsymbol{\mu}\) shift of \(\Lagr\) or \(\dd\mathbf Y\) shift of \(\Thetaform\) is imposed. With this convention, \(\SH\) is the entropy extracted from the black hole first law, while \(\Sone\) and \(\DeltaS\) are computed from the corresponding entropy-sector pair \(\QS\) and \(\ThetaS\), after the ordinary non-gravitational work sector has been separated. They are not introduced by an ambiguity transformation.

For a stationary solution and the horizon Killing generator \(\xi^a\), \(\Lie_{\xi}\Phi=0\). If the background is on shell and the variation satisfies the linearized equations of motion, then
\[
\omegaform[\Phi;\delta\Phi,\Lie_\xi\Phi]=0,
\qquad
\dd\kxi=0 .
\]
Integrating \(\dd\kxi=0\) over a spacelike hypersurface \(\Sigma\) with outer boundary \(S_r\) and inner boundary \(\Bhor\) gives
\begin{equation}
	\label{eq:FirstLawGen}
	\begin{aligned}
		0
		&=
		\oint_{S_r}
		\Bigl(
		\delta\mathbf Q_{\xi}
		-\xi\cdot\Thetaform
		\Bigr)-
		\oint_{\Bhor}
		\Bigl(
		\delta\mathbf Q_{\xi}
		-\xi\cdot\Thetaform
		\Bigr).
	\end{aligned}
\end{equation}
This is the covariant phase space identity from which the black hole first law is read. The outer boundary \(S_r\) is a large radius surface; the asymptotic charge variation is obtained only after taking \(r\to\infty\). The required integrability condition is
\begin{equation}
	\label{eq:asymptotic-integrability}
	\lim_{r\to\infty}
	\oint_{S_r}
	\Bigl(
	\delta\mathbf Q_{\xi}
	-\xi\cdot\Thetaform
	\Bigr)
	=
	\delta H_{\xi}^{(\infty)}
\end{equation}
up to ordinary work terms. We assume this integrability when extracting the entropy from the horizon surface charge variation. If the asymptotic charge variation is nonintegrable, Eq.~\eqref{eq:FirstLawGen} still relates the two boundary variations, but the horizon variation alone does not define a global entropy correction.

Before identifying the entropy part of the horizon charge, we separate the
ordinary non-gravitational boundary one-form. In the fixed representative
obtained by varying the displayed Lagrangian, write
\begin{equation}
	\label{eq:S-m-split}
	\Qxi=\QS+\mathbf Q^{\rm m}_\xi,\qquad
	\Thetaform=\ThetaS+\mathbf\Theta_{\rm m}.
\end{equation}
The label ``m'' denotes boundary terms that give ordinary matter or gauge work,
such as Maxwell, or vector kinetic contributions carrying an
independent charge or a fixed background variation. These terms are not counted
as entropy corrections. The remaining entropy-sector surface form is
\begin{equation}
	\label{eq:k-S-def}
	\kS
	=
	\delta\QS-\xi\cdot\ThetaS .
\end{equation}
The matter/gauge work one-form is then
\begin{equation}
	\label{eq:Wm-def}
	\Wm
	=
	\oint_{\Bhor}
	\Bigl(
	\delta\mathbf Q^{\rm m}_\xi
	-\xi\cdot\mathbf\Theta_{\rm m}
	\Bigr),
\end{equation}
with the understanding that on a fixed-background or fixed-charge slice this
one-form may vanish. This split is only a bookkeeping of the directly varied
representative; it is not a JKM transformation.

The horizon surface charge variation used below for the entropy sector is
\begin{equation}
	\label{eq:horizon-surface charge}
	\oint_{\Bhor}\kS
	=
	\oint_{\Bhor}
	\Bigl(
	\delta \QS
	- \xi\cdot\ThetaS
	\Bigr).
\end{equation}
If the chosen solution family is enlarged so that conserved charges or other integration constants are varied, the black hole first law generally contains thermodynamic work terms,
\begin{equation}
	\label{eq:work-one-form}
	\mathcal W_{\rm th}
	=
	\Wm+\Wpar,
	\qquad
	\Wpar=
	\Phi_{\mathcal Q}\,\delta\mathcal Q
	+\sum_I \Phi_I\,\delta\eta_I .
\end{equation}
Here \(\mathcal Q\) is a conserved gauge or electric charge not already included in \(\Wm\), and \(\eta_I\) denotes any additional black hole parameter varied in the chosen solution space, such as a coupling, hair parameter, or branch integration constant. The conjugate potentials are fixed by the full thermodynamic first law. These terms are work terms, not entropy terms, so the entropy variation is defined after subtracting them from the total horizon charge, or equivalently from the entropy-sector charge:
\begin{equation}
	\label{eq:entropy-definition}
	\frac{\kappa}{2\pi}\,\delta \SH
	=\oint_{\Bhor}\kxi-\mathcal W_{\rm th}
	=\oint_{\Bhor}\kS-\Wpar .
\end{equation}
We denote the corresponding fixed parameter variation by \(\widehat{\delta}\):
\[
\widehat{\delta}\mathcal Q=0,
\qquad
\widehat{\delta}\eta_I=0,
\qquad
\Wm[\widehat{\delta}]=0,
\qquad
\mathcal W_{\rm th}[\widehat{\delta}]=0 .
\]
This restricted variation is only a computational projection of the full thermodynamic first law one-form. It leaves the thermodynamic phase space unchanged; only the variations of the work parameters are set to zero. On this slice \((\kappa/2\pi)\widehat{\delta}\SH=\oint_{\Bhor}\kS\big|_{\delta=\widehat{\delta}}\).

\section{Horizon decomposition and thermodynamic first law}
\label{sec:horizon-decomposition}

\subsection{Stationary horizon setup and entropy decomposition}

The symbol \(\delta\QS\) denotes the variation of the entropy-sector charge form at fixed vector label \(\xi^a\). If the label is varied as well, the total variation contains an additional term because the Noether charge is a linear differential expression in \(\xi^a\):
\begin{equation}
	\label{eq:delta-Qxi-horizon}
	\begin{aligned}
		\delta\!\left(\oint_{\Bhor}\QS\right)
		&=
		\oint_{\Bhor}\delta\QS
		+\oint_{\Bhor}\QdS ,
		\\
		\oint_{\Bhor}\delta\QS
		&=
		\delta\!\left(\oint_{\Bhor}\QS\right)
		-\oint_{\Bhor}\QdS .
	\end{aligned}
\end{equation}
Here \(\QdS\) denotes the same local entropy-sector Noether charge form with \(\xi^a\) replaced by \(\delta\xi^a\).

We now specialize to static, spherically symmetric black holes with Killing horizon generator \(\xi^a=(\partial_t)^a\). Rotating families, for which one would also need the rotational Killing field \((\partial_\phi)^a\) and the corresponding angular momentum term, are not considered here. In the solution families studied below, the same asymptotic time translation Killing field is used for every member, so
\begin{equation}
	\delta\xi^a=0,
	\qquad
	\delta\xi_a
	=
	\delta(g_{ab}\xi^b)
	=
	\xi^b\delta g_{ab}.
\end{equation}
Thus \(\QdS\) vanishes in the final static formulas, but \(\delta\xi_a\) need not vanish. The variation of \(\nabla_a\xi_b\), the variation of the surface gravity, and the entropy-sector presymplectic potential contribution \(-\xi\cdot\ThetaS\) can therefore remain nontrivial. The horizon temperature is denoted by \(T=\kappa/(2\pi)\). After the work terms are subtracted, the remaining entropy variation is assumed to be integrable on the chosen solution family, so \(\SH\) is the entropy entering the first law for that family rather than a universal local functional on arbitrary \(\Bhor\).

For the representative obtained directly from the displayed Lagrangian, we use the standard Iyer--Wald decomposition of the entropy-sector Noether charge. For theories whose entropy-sector Noether charge admits the usual split into an \(X\)-part and a \(W\)-part, one may write schematically \cite{Wald:1993nt,Iyer:1994ys}
\begin{equation}
	\label{eq:Q-XW-split}
	\QS
	=
	\mathbf W_a\,\xi^a
	+ \mathbf X^{ab}\nabla_{[a}\xi_{b]} .
\end{equation}
The \(X\)-term gives the standard Wald entropy,
\begin{equation}
	\label{eq:Sw-def}
	\SW
	=
	-2\pi\oint_{\Bhor} E_R^{abcd}\,\epsilon_{ab}\epsilon_{cd}.
\end{equation}
Away from the regular bifurcation surface setting, the full horizon integral of \(\QS\) need not be exhausted by the Wald term. We define the entropy read from the entropy-sector horizon Noether charge and its non-Wald part by
\begin{align}
	\label{eq:Sq-shift}
	\Sone
	&
	=
	\frac{2\pi}{\kappa}\oint_{\Bhor} \mathbf W_a\,\xi^a,
	\\
	\label{eq:S1-W-def}
	\SQ
	&=
	\frac{2\pi}{\kappa}\oint_{\Bhor} \QS
	=
	\SW+\Sone
	.
\end{align}
The second line follows after the \(X\)-part that gives \(\SW\) has been separated. The term \(\mathbf W_a\xi^a\) may contain contributions involving \(\nabla_dE_R^{abcd}\), but is not identified with that term alone. 

The remaining integrable part of the horizon surface charge variation is denoted by \(\DeltaS\). Thus the entropy entering the first law on the chosen solution family is
\begin{equation}
	\label{eq:S-main}
	\SH=\SQ+\DeltaS=\SW+\Sone+\DeltaS .
\end{equation}
In regular bifurcate situations both corrections vanish, and \(\SH=\SW\).

Eq.~\eqref{eq:FirstLawGen} fixes the relative sign between the outer and horizon boundaries. With the fixed-generator choice already imposed, the horizon surface charge variation Eq.~\eqref{eq:horizon-surface charge} becomes
\begin{equation}
	\label{eq:horizon-charge-SQ-split}
	\oint_{\Bhor}\kS
	=
	\delta\!\left(\frac{\kappa\SQ}{2\pi}\right)
	-\oint_{\Bhor}\xi\cdot\ThetaS .
\end{equation}
Combining this identity with the entropy definition Eqs.~\eqref{eq:entropy-definition} and~\eqref{eq:S-main} gives
\begin{equation}
	\label{eq:master-entropy}
	\begin{aligned}
		\frac{2\pi}{\kappa}
		\left[
		\left(-\oint_{\Bhor}\xi\cdot\ThetaS\right)
		-\Wpar
		\right]
		=
		-\frac{\delta\kappa}{\kappa}\,\SQ
		+\delta\!\DeltaS .
	\end{aligned}
\end{equation}

For the calculations below it is sufficient to use the fixed parameter direction \(\widehat{\delta}\). Then
\[
\widehat{\delta}\mathcal Q=0,
\qquad
\widehat{\delta}\eta_I=0,
\qquad
\Wpar[\widehat{\delta}]=0,
\qquad
\mathcal W_{\rm th}[\widehat{\delta}]=0,
\]
and Eq.~\eqref{eq:master-entropy} reduces to
\begin{equation}
	\label{eq:DeltaS-restricted-general}
	\widehat{\delta}\!\DeltaS
	=
	\frac{2\pi}{\kappa}
	\left(-\oint_{\Bhor}\xi\cdot\ThetaS\right)_{\delta=\widehat{\delta}}
	+\frac{\widehat{\delta}\kappa}{\kappa}\,\SQ .
\end{equation}

When the regular bifurcation surface assumptions of the Iyer--Wald construction hold, both corrections vanish:
\begin{equation}
	\Sone=0,\qquad \DeltaS=0.
\end{equation}
Thus \(\SQ=\SH=\SW\).

For the static families studied below, Eq.~\eqref{eq:master-entropy} gives
\begin{equation}
	\label{eq:DeltaS-from-theta}
	\delta\!\DeltaS
	=
	\frac{2\pi}{\kappa}
	\left[
	\left(-\oint_{\Bhor} \xi\cdot\ThetaS\right)
	-\Wpar
	\right]
	+\frac{\delta\kappa}{\kappa}\,\SQ,
\end{equation}
On the restricted slice this becomes
\begin{equation}
	\label{eq:DeltaS-static-restricted}
	\widehat{\delta}\!\DeltaS
	=
	\frac{2\pi}{\kappa}
	\left(-\oint_{\Bhor} \xi\cdot\ThetaS\right)_{\delta=\widehat{\delta}}
	+\frac{\widehat{\delta}\kappa}{\kappa}\,\SQ.
\end{equation}

\subsection[Decomposition of minus xi dot Theta S]{Decomposition of \texorpdfstring{\(-\xi\cdot\ThetaS\)}{-xi dot Theta S}}

We now state the local horizon decomposition used in the examples. The
assumptions are the following. First, the covariant phase space representative
is the one obtained by directly varying the displayed Lagrangian; no additional
JKM counterterms are added. Second, the horizon generator is field independent,
\(\delta\xi^a=0\), and the horizon surface charge variation is evaluated either on a regular
bifurcation surface or as the limit of a stretched surface in coordinates
adapted to the Killing horizon. Third, the sign of the horizon integral is the
inner boundary sign used in Eq.~\eqref{eq:FirstLawGen}, and the variations preserve the stationary
horizon gauge and the chosen static, spherically symmetric solution family.
Under these assumptions, the part of \(\ThetaS\) relevant to the entropy
variation admits the following decomposition.

We consider theories of the form~\eqref{eq:lagrangian-form}.
After the non-gravitational work one-form has been separated, the relevant
entropy-sector presymplectic potential contribution on a stationary Killing
horizon can be written as
\begin{equation}
	\begin{aligned}
		(\Theta_{\rm S})_{bcd}
		&=
		\epsilon_{abcd}\,(\theta_X^a+\theta_W^a),
		\\
		\theta_X^a
		&=
		2E_R^{abcd}\nabla_d\delta g_{bc},
		\\
		\theta_W^a
		&=
		-2(\nabla_dE_R^{abcd})\delta g_{bc} .
	\end{aligned}
	\label{eq:theta-XW-def}
\end{equation}
Here \(\theta_X^a\) is the part of the presymplectic potential associated with the \(X^{ab}\nabla_{[a}\xi_{b]}\) term in Eq.~\eqref{eq:Q-XW-split}.
The term \(-2(\nabla_dE_R^{rbcd})\delta g_{bc}\) in \(\theta_W^r\) corresponds to the curvature derivative contribution \(-2\xi_c\nabla_d E_R^{mncd}\) in the Noether charge form, which is contained in \(\mathbf W_a\xi^a\).

The split \(\theta^a=\theta_X^a+\theta_W^a\) is made only after the Lagrangian representative and presymplectic potential have been fixed. A different JKM representative can move terms between the local \(X\)- and \(W\)-part densities. We therefore fix the representative obtained by directly varying the displayed Lagrangian, and all horizon integrals below are evaluated in this representative on the chosen solution family.

Choose Gaussian null coordinates \((v,r,x^A)\) adapted to the horizon
\cite{Racz:1992bp,Chrusciel:2012jk,Kunduri:2013gce}, so that
\begin{equation}
	\label{eq:gnc-horizon}
	\xi^a=(\partial_v)^a,
	\qquad
	\Bhor=\{r=0,\ v=\text{const}\},
\end{equation}
with radial component
\begin{equation}
	\label{eq:gnc-metric}
	\begin{aligned}
		ds^2
		&=
		-2\kappa r\,dv^2+2\,dv\,dr+2r\,\alpha_A\,dv\,dx^A
		\\
		&\quad
		+q_{AB}(r,x)\,dx^A dx^B+O(r^2).
	\end{aligned}
\end{equation}
Let \(\tilde{\boldsymbol\epsilon}\) be the \((D-2)\)-form on \(\Bhor\) with the inner-boundary orientation,
\(\tilde{\epsilon}_{A_1\cdots A_{D-2}}\equiv
\epsilon_{r v A_1\cdots A_{D-2}}\).
Then
\begin{equation}
	\label{eq:xi-theta-component}
	(\xi\cdot\ThetaS)_{A_1\cdots A_{D-2}}
	=
	\tilde{\epsilon}_{A_1\cdots A_{D-2}}\theta^r,
\end{equation}
so that
\begin{equation}
	\label{eq:xi-theta-r-component}
	-\oint_{\Bhor} \xi\cdot\ThetaS
	=
	-\oint_{\Bhor} \tilde{\boldsymbol\epsilon}\,(\theta_X^r+\theta_W^r).
\end{equation}

We use the notation
\[
E_R^{r(bc)d}=\frac12\bigl(E_R^{rbcd}+E_R^{rcbd}\bigr).
\]
Eq.~\eqref{eq:xi-theta-r-component} then gives
\begin{align}
	\theta_X^r
	&=
	2E_R^{r(vv)d}\nabla_d\delta g_{vv}
	\notag\\
	&\quad
	+4E_R^{r(vr)d}\nabla_d\delta g_{vr}
	\notag\\
	&\quad
	+4E_R^{r(vA)d}\nabla_d\delta g_{vA}
	\notag\\
	&\quad
	+4E_R^{r(rA)d}\nabla_d\delta g_{rA}
	+2E_R^{r(AB)d}\nabla_d\delta q_{AB}.
	\label{eq:thetaX-decomp}
\end{align}

For variations preserving Gaussian null gauge and keeping the horizon at \(r=0\), one has
\begin{equation}
	\label{eq:delta-kappa-gnc}
	\delta\kappa=-\frac12\,\nabla_r\delta g_{vv}\big|_{\Bhor}.
\end{equation}
With this sign convention, the Wald entropy has the component form
\begin{equation}
	\label{eq:Sw-component-gnc}
	\SW
	=
	-8\pi\oint_{\Bhor}\tilde{\boldsymbol\epsilon}\,E_R^{r(vv)r}.
\end{equation}
The \(d=r\) piece of the first term in Eq.~\eqref{eq:thetaX-decomp} therefore reproduces the standard Wald contribution:
\begin{equation}
	\label{eq:thetaX-integral}
	-\oint_{\Bhor} \tilde{\boldsymbol\epsilon}\,\theta_X^r
	=
	-\frac{\SW}{2\pi}\,\delta\kappa+\mathcal R_X,
\end{equation}
with
\begin{align}
	\mathcal{R}_X
	&= 
	-2\oint_{\Bhor} \tilde{\boldsymbol\epsilon}\,
	\Big[
	E_R^{r(vv)A}\nabla_A\delta g_{vv}
	\notag\\
	&\qquad
	+2E_R^{r(vr)d}\nabla_d\delta g_{vr}
	\notag\\
	&\qquad
	+2E_R^{r(vA)d}\nabla_d\delta g_{vA}
	\notag\\
	&\qquad
	+2E_R^{r(rA)d}\nabla_d\delta g_{rA}
	+E_R^{r(AB)d}\nabla_d\delta q_{AB}
	\Big].
	\label{eq:Rx-def}
\end{align}
For the static, spherically symmetric solution families considered below, the
allowed variations preserve the ansatz. Thus
\[
\delta g_{vr}=\delta g_{vA}=\delta g_{rA}=0
\]
near \(\Bhor\). The angular metric has the form
\(q_{AB}=R^2(r)\gamma_{AB}\), with \(\gamma_{AB}\) the unit round metric.
Its variation therefore has no \(v\) dependence or angular dependence:
\(\nabla_v\delta q_{AB}=0\) and \(\nabla_C\delta q_{AB}=0\) on \(\Bhor\).
Consequently all terms in Eq.~\eqref{eq:Rx-def} vanish except the possible radial
derivative of \(\delta q_{AB}\),
\begin{equation}
	\label{eq:Rx-static-spherical}
	\mathcal R_X
	=
	-2\oint_{\Bhor}\tilde{\boldsymbol\epsilon}\,
	E_R^{r(AB)r}\nabla_r\delta q_{AB}.
\end{equation}
For the Einstein-Hilbert Lagrangian, the only nonzero component in the unreduced gauge-preserving expression has \(d=v\), while the stationary variations used here satisfy
\(\nabla_v\delta q_{AB}=0\) on \(\Bhor\). Thus \(\mathcal R_X=0\) in general
relativity. 

Similarly, the \(W\)-part can be written as
\begin{equation}
	\label{eq:thetaW-integral}
	\mathcal R_W
	=\frac{\Sone}{2\pi}\,\delta\kappa-\oint_{\Bhor} \tilde{\boldsymbol\epsilon}\,\theta_W^r,
\end{equation}
which defines \(\mathcal R_W\) in the same way that Eq.~\eqref{eq:thetaX-integral} defines \(\mathcal R_X\).
For the Einstein-Hilbert Lagrangian, \(\nabla_dE_R^{abcd}=0\), so
\(-\oint_{\Bhor}\tilde{\boldsymbol\epsilon}\,\theta_W^r=0\).

The term proportional to \(\delta\kappa\) is fixed by the non-Wald Noether charge correction \(\Sone\), while \(\mathcal R_W\) contains the remainder of the presymplectic potential contribution associated with \(\mathbf W_a\xi^a\).
It should not be identified with the single curvature derivative term \(-2(\nabla E_R)\delta g\).
Combining the two parts gives
\begin{equation}
	\label{eq:xi-theta-decomp-main}
	\begin{aligned}
		-\oint_{\Bhor} \xi\cdot\ThetaS
		&=
		-\frac{\SW+\Sone}{2\pi}\,\delta\kappa
		+\mathcal R_X+\mathcal R_W
		\\
		&=
		-\frac{\SQ}{2\pi}\,\delta\kappa
		+\mathcal R_X+\mathcal R_W .
	\end{aligned}
\end{equation}
Together with Eq.~\eqref{eq:DeltaS-from-theta}, this yields the compact entropy identity
\begin{equation}
	\label{eq:Rx-Rw-DeltaS}
	\frac{2\pi}{\kappa}
	\left(\mathcal R_X+\mathcal R_W-\Wpar\right)
	=
	\delta\!\DeltaS
\end{equation}
for a full variation on an enlarged thermodynamic phase space. On the restricted slice we write
\begin{equation}
	\label{eq:restricted-remainders-def}
	\RhX=\mathcal R_X\big|_{\delta=\widehat{\delta}},
	\qquad
	\RhW=\mathcal R_W\big|_{\delta=\widehat{\delta}}.
\end{equation}
Thus
\begin{equation}
	\label{eq:restricted-horizon-decomposition}
	\begin{aligned}
		\left(-\oint_{\Bhor} \xi\cdot\ThetaS\right)_{\delta=\widehat{\delta}}
		&=
		-\frac{\SQ}{2\pi}\,\widehat{\delta}\kappa
		+\RhX+\RhW,
		\\
		\widehat{\delta}\!\DeltaS
		&=
		\frac{2\pi}{\kappa}\left(\RhX+\RhW\right).
	\end{aligned}
\end{equation}
They should be distinguished from the full \(\mathcal R_X,\mathcal R_W\), which may contain additional pieces proportional to variations of parameters such as \(\delta\eta_I\).

\subsection{Entropy criterion}

The preceding calculation gives the following test for the entropy that enters the first law. The bifurcation surface reduction applies when the metric and nonmetric fields entering the entropy-sector couplings can be smoothly extended to the regular bifurcation surface. Then the \(\mathbf W_a\xi^a\) charge term and the remaining entropy-sector presymplectic potential contribution vanish in the usual Iyer--Wald way, and \(\SH=\SW\). If some background field or curvature-coupled combination is singular in regular horizon coordinates, finite pieces in \(\Sone\) or \(\DeltaS\) can survive. The quantity \(\DeltaS\) is local only if the solution-space integral can be rewritten as a local horizon functional.

Consider a stationary family of black hole solutions in a diffeomorphism-invariant theory with \(\Lagr=\boldsymbol{\epsilon}\,L(g_{ab},R_{abcd},\phi^I,\nabla\phi^I)\). Evaluate the entropy-sector horizon surface charge variation on the horizon cross section, either directly at a regular bifurcation surface when this is available or by the corresponding stretched horizon limit. Assume:
\begin{enumerate}
	\item the family admits a Killing horizon generator \(\xi^a=(\partial_t)^a\);
	\item the entropy one-form obtained from the equality between \(\delta H_{\xi}^{(\infty)}\) and the horizon variation in Eq.~\eqref{eq:horizon-surface charge} is integrable on the chosen solution family.
\end{enumerate}
Then
\begin{equation}
	\label{eq:criterion-entropy-decomposition}
	\SH=\SW+\Sone+\DeltaS,
\end{equation}
with the three terms defined in Eqs.~\eqref{eq:Sw-def}, \eqref{eq:S1-W-def}, and \eqref{eq:DeltaS-from-theta}. Here \(\Sone\) and \(\DeltaS\) are defined only after the split \eqref{eq:S-m-split}. For the restricted variation \(\widehat{\delta}\), the remainders are \(\RhX,\RhW\), with
\[
\widehat{\delta}\!\DeltaS
=
\frac{2\pi}{\kappa}\left(\RhX+\RhW\right).
\]
The departure from the Wald entropy is then controlled by the Noether charge contribution \(\oint_{\Bhor}\mathbf W_a\xi^a\) and by the remainder \(\RhX+\RhW\):
\begin{enumerate}
	\item if \(\oint_{\Bhor}\mathbf W_a\xi^a=0\) and \(\RhX+\RhW=0\), then \(\Sone=0\), \(\DeltaS=0\), and \(\SH=\SW\);
	\item if \(\oint_{\Bhor}\mathbf W_a\xi^a=0\) but \(\RhX+\RhW\neq0\), then the correction is entirely in the remaining part of \(-\oint_{\Bhor}\xi\cdot\ThetaS\) on the restricted slice;
	\item if \(\oint_{\Bhor}\mathbf W_a\xi^a\neq0\) but \(\RhX+\RhW=0\), then the correction is entirely in the non-Wald part of the entropy-sector horizon Noether charge;
	\item if \(\oint_{\Bhor}\mathbf W_a\xi^a\neq0\) and \(\RhX+\RhW\neq0\), then both \(\Sone\) and \(\DeltaS\) must be kept in \(\SH\).
\end{enumerate}

\section{Illustrative models}
\label{sec:models}

We now apply the preceding decomposition to several static, spherically symmetric black hole solutions. The examples are chosen to separate the different sources in Eq.~\eqref{eq:criterion-entropy-decomposition}: a regular nonminimally coupled tensor background for which the additional terms vanish, vector-tensor branches where the remaining \(-\oint_{\Bhor}\xi\cdot\ThetaS\) contribution is essential, and higher-curvature examples with a Weyl vector where the non-Wald entropy-sector Noether charge term and the remaining horizon contribution both enter.

Throughout this section we set \(G=1\) in the thermodynamic quantities.

In each example the displayed \(\Thetaform\) and \(\Qxi\) are first split according to Eq.~\eqref{eq:S-m-split}. We then compute \(\SW\) and \(\Sone\), determine \(\DeltaS\) from
\begin{equation}
	\label{eq:model-restricted-DeltaS}
	\frac{2\pi}{\kappa}\left(\RhX+\RhW\right)
	=
	\widehat{\delta}\!\DeltaS ,
\end{equation}
assemble \(\SH=\SW+\Sone+\DeltaS\), and compare the result with the thermodynamic first law.

\subsection{Gravity coupled with a background	Kalb-Ramond field}

As a concrete matching example, we consider the Einstein--Hilbert action
nonminimally coupled to a self-interacting Kalb--Ramond
two-form~\cite{Kalb:1974yc,Altschul:2009ae,Hernaski:2016dyk}. To avoid confusion with
the horizon surface gravity \(\kappa\), we denote the gravitational coupling in
the action by \(\kappa_g=8\pi G\). The model takes the form~\cite{Altschul:2009ae}
\begin{align}
	S
	&=
	\int d^4x\,\sqrt{-g}\Big[
	\frac{1}{2\kappa_g}(R-2\Lambda)
	-\frac{1}{12}H^{\mu\nu\rho}H_{\mu\nu\rho}
	\notag\\
	&\qquad
	-V(B^{\mu\nu}B_{\mu\nu}\pm b^2)
	+\frac{1}{2\kappa_g}\Big(
	\zeta_1 B^{\mu\nu}B_{\mu\nu}R
	\notag\\
	&\qquad\qquad
	+\zeta_2 B^{\rho\mu}B^\nu{}_\mu R_{\rho\nu}
	\Big)
	\Big],
\end{align}
where
\begin{equation}
	H_{\mu\nu\rho}
	=
	\partial_\mu B_{\nu\rho}
	+\partial_\nu B_{\rho\mu}
	+\partial_\rho B_{\mu\nu}.
\end{equation}
Here \(B_{\mu\nu}\) is the Kalb--Ramond tensor field and
\(H_{\mu\nu\rho}\) is its field strength. The potential is written as a
function of
\(X_{\rm KR}=B^{\mu\nu}B_{\mu\nu}\pm b^2\). The constant \(b^2\) controls the
nonzero tensor vacuum expectation value through the vacuum condition
\(X_{\rm KR}=0\); for example, a quadratic potential
\(V(X_{\rm KR})=\lambda X_{\rm KR}^2/2\) has \(V=V'=0\) on this vacuum branch.
The constants \(\zeta_1\) and \(\zeta_2\) are nonminimal tensor couplings.
Static and spherically symmetric black hole solutions in this theory were obtained in Refs.~\cite{Yang:2023wtu,Duan:2023gng,Liu:2024oas,Liu:2025fxj}. We consider the Schwarzschild-like solution with \(\zeta_1=0\)~\cite{Yang:2023wtu}
\begin{equation}
	\begin{aligned}
		ds^2
		&=
		-A(r)\,dt^2
		+S(r)\,dr^2
		+r^2 d\Omega_2^2,
	\end{aligned}
\end{equation}
with
\begin{equation}
	\begin{aligned}
		A(r)&=\frac{1}{1-\ell}-\frac{2M}{r},\\
		S(r)&=\frac{1}{A(r)},\\
		B_{tr}&=-B_{rt}=\frac{|b|}{\sqrt{2}}.
	\end{aligned}
\end{equation}
Since \(b\) is a constant background value and \(\ell=\zeta_2 b^2\), the effective parameter \(\ell\) is also constant. We treat \(b\) and \(\ell\) as fixed background data, so \(\widehat{\delta}\ell=0\).
Denoting by \(r_h\) the horizon radius determined by \(A(r_h)=0\), the horizon analysis for the Killing generator \(\xi\) gives the directly varied presymplectic potential and Noether charge~\cite{Liu:2025fxj,Xiao:2025icr}
\begin{align}
	\Thetaform[\Phi,\delta\Phi]_{bcd}
	&=
	\Bigl(
	2E_R^{aefh}\nabla_h\delta g_{ef}
	-2\left(\nabla_hE_R^{aefh}\right)\delta g_{ef}
	\notag\\
	&\quad
	-\frac12 H^{aef}\delta B_{ef}
	\Bigr)\varepsilon_{abcd},
	\\
	(\Qxi)_{cd}
	&=
	\Bigl(
	-E_R^{abef}\nabla_e\xi_f
	-2\xi_e\nabla_fE_R^{abef}
	\notag\\
	&\quad
	+\frac12 H^{abe}B_{ef}\xi^f
	\Bigr)\varepsilon_{abcd}.
\end{align}
Here
\begin{align}
	E_R^{abcd}
	&=
	\frac{1}{2\kappa_g}X^{abcd}
	+\frac{\zeta_2}{2\kappa_g}B^{\mu\rho}B^\nu{}_\rho
	{Y_{\mu\nu}}^{abcd},
	\notag\\
	X^{abcd}
	&=
	g^{a[c}g^{d]b},
	\notag\\
	{Y_{\mu\nu}}^{abcd}
	&=
	\frac12\left(
	{g_{(\mu}}^a {g_{\nu)}}^{[c}g^{d]b}
	-{g_{(\mu}}^b {g_{\nu)}}^{[c}g^{d]a}
	\right).
\end{align}
Since \(S=1/A\), the surface gravity is
\begin{equation}
	\kappa
	=
	\frac{A'(r_h)}{2}
	=
	\frac{1}{2(1-\ell)r_h}.
\end{equation}
The Wald entropy and the entropy-sector \(\mathbf W_a\xi^a\) charge contribution are
\begin{equation}
	\begin{aligned}
		\SW=(1+\ell)\pi r_h^2,\qquad
		\oint_{\Bhor} \mathbf W_a\,\xi^a=0,
		\qquad
		\Sone=0 .
	\end{aligned}
\end{equation}
For this black hole family, the horizon decomposition is particularly simple:
\begin{equation}
	\begin{aligned}
		\RhX=0,\qquad \RhW=0.
	\end{aligned}
\end{equation}
The restricted identity therefore gives
\begin{equation}
	\widehat{\delta}\!\DeltaS=0.
\end{equation}
Choosing the integration constant so that the Einstein limit is continuous,
\begin{equation}
	\DeltaS=0,
\end{equation}
and combining these contributions gives the entropy entering the first law,
\begin{equation}
	\SH=\SW+\Sone+\DeltaS=\SW .
\end{equation}
Only after these contributions are combined do we compare with the thermodynamic first law. The asymptotic surface charge gives
\begin{equation}
	E=(1+\ell)M,
\end{equation}
and on the restricted slice
\begin{equation}
	\widehat{\delta}E
	=
	T\widehat{\delta}\SH
	=
	T\widehat{\delta}\SW .
\end{equation}
Thus this nonminimally coupled Kalb--Ramond black hole has the same Wald entropy and first-law entropy. The coupling rescales the Wald entropy, but after the ordinary Kalb--Ramond boundary term is separated the tensor field is sufficiently regular for the entropy-sector \(\mathbf W_a\xi^a\) and \(\xi\cdot\ThetaS\) contributions to vanish.

\subsection{Bumblebee gravity}

Bumblebee gravity is a vector-tensor realization of spontaneous Lorentz
symmetry breaking in the gravitational sector
\cite{Kostelecky:1988zi,Bluhm:2004ep,Bailey:2006fd,Kostelecky:2003fs}. A
commonly used four-dimensional action is~\cite{Kostelecky:2003fs}
\begin{align}
	S
	&=
	\int d^4x\sqrt{-g}\Big[
	\frac{1}{2\kappa_g}(R-2\Lambda)
	+\frac{\xi_B}{2\kappa_g}B^{\mu}B^{\nu}R_{\mu\nu}
	\notag\\
	&\qquad
	-\frac{1}{4}\mathcal B_{\mu\nu}\mathcal B^{\mu\nu}
	-V(B^\mu B_\mu\pm b^2)
	\Big],
	\label{eq:bumblebee-action}
\end{align}
where \(B_\mu\) is the bumblebee vector field and
\(\mathcal B_{\mu\nu}=\nabla_\mu B_\nu-\nabla_\nu B_\mu\) is its field strength. The
nonminimal coupling \(B^\mu B^\nu R_{\mu\nu}\) feeds the preferred direction
back into the metric equations. The potential is usually taken as a function of
\(X_B=B^\mu B_\mu\pm b^2\). On the vacuum branch, the constraint \(X_B=0\) imposes
\(B^\mu B_\mu=\mp b^2\), so that the vector field acquires
\(\langle B_\mu\rangle=b_\mu\) with
\(b^\mu b_\mu=\mp b^2=\mathrm{const}\). The resulting vacuum configuration can
be spacelike, timelike, or lightlike. Exact black hole solutions in bumblebee
gravity have been constructed~\cite{Casana:2017jkc,Maluf:2020kgf,Liu:2024axg,Liu:2025oho,Zhu:2025fiy}. For the
solution ansatz below, we choose the quadratic potential \(V(X_B)=\lambda X_B^2/2\).
On the vacuum branch \(X_B=0\), this gives \(V=V'=0\). We use the static metric ansatz
\begin{equation}
	\begin{aligned}
		ds^2
		&=
		-A(r)\,dt^2
		+S(r)\,dr^2
		+r^2 d\Omega_2^2,
	\end{aligned}
\end{equation}
and take the background bumblebee profile to be
\begin{equation}
	B_{\mu}=b_\mu=\left(b_t(r),b_r(r),0,0\right).
	\label{eq:bumblebee-bmu-ansatz}
\end{equation}

For the Killing generator \(\xi^a\), the directly varied horizon presymplectic potential and Noether charge take the common form
\begin{align}
	\Thetaform[\Phi,\delta\Phi]_{bcd}
	&=
	\Bigl(
	2{E_R}^{aefh}\nabla_{h}\delta g_{ef}
	- 2\left(\nabla_{h}{E_R}^{aefh}\right)\delta g_{ef}
	\notag\\
	&\quad
	- \mathcal B^{ae}\delta B_{e}
	\Bigr)\varepsilon_{abcd},
	\notag\\
	(\Qxi)_{cd}
	&=
	\Bigl(
	- {E_R}^{abef}\nabla_{e}\xi_{f}
	- 2\xi_{e}\nabla_{f}{E_R}^{abef}
	\notag\\
	&\quad
	- \frac{1}{2}\mathcal B^{ab}B_{f}\xi^{f}
	\Bigr)\varepsilon_{abcd},
	\label{eq:bumblebee-theta-K}
\end{align}

with
\begin{align}
	{E_R}^{abcd}
	&=
	\frac{1}{2\kappa_g}\left(
	X^{abcd}+\xi_B B^{e}B^{f}{Y_{ef}}^{abcd}
	\right),
	\label{eq:bumblebee-ER}
	\\
	X^{abcd}
	&=
	g^{a[c}g^{d]b},
	\\
	{Y_{ef}}^{abcd}
	&=
	\frac{1}{2}\bigl(
	{g_{(e}}^a{g_{f)}}^{[c}g^{d]b}
	- {g_{(e}}^b{g_{f)}}^{[c}g^{d]a}
	\bigr).
\end{align}
Here \(\xi_B\) denotes the nonminimal bumblebee coupling constant to avoid confusion with the Killing field \(\xi^a\). The two static branches considered below differ in the background profile and in how the charge and presymplectic potential terms evaluate at the horizon.
In contrast with the Kalb--Ramond branch, the bumblebee one-form can be singular in a regular horizon frame. Then the standard bifurcation surface argument does not automatically remove the \(\mathbf W_a\xi^a\) charge contribution or the associated presymplectic potential terms.

\paragraph{Spacelike branch.}
For the spacelike vacuum expectation value \(B_\mu B^\mu=b^2\), we have~\cite{Casana:2017jkc}
\begin{equation}
	b_r(r)=b\,\sqrt{S(r)},\qquad
	b_t(r)=0
	\label{btr}
\end{equation}
with
\begin{equation}
	A(r)=1-\frac{2M}{r},
	\qquad
	S(r)=\frac{1+\ell_1}{A(r)},
	\qquad
	\ell_1=\xi_B b^2.
\end{equation}
The parameter \(b\) fixes the chosen spacelike bumblebee background, so \(\ell_1\) is held fixed and \(\widehat{\delta}\ell_1=0\).
Because \(S(r)=(1+\ell_1)A(r)^{-1}\) at the horizon, the radial one-form component is not regular in a freely falling horizon frame.
The surface gravity is
\begin{equation}
	\kappa
	=
	\left.
	\frac{A'(r)}{2\sqrt{1+\ell_1}}
	\right|_{r=r_h}
	=
	\frac{1}{2r_h\sqrt{1+\ell_1}} .
\end{equation}
The entropy-sector horizon Noether charge integral agrees with the Wald entropy~\cite{An:2024fzf,Liu:2025oho},
\begin{equation}
	\SW=\SQ=\left(1+\frac{\ell_1}{2}\right)\pi r_h^2 .
\end{equation}
Thus \(\Sone\) vanishes, $\Sone=0$.

The remaining correction comes only from \(-\oint_{\Bhor}\xi\cdot\ThetaS\). In this branch
\begin{equation}
	\RhX=0,
	\qquad
	\frac{2\pi}{\kappa}\RhW
	=
	\widehat{\delta}\!\DeltaS
	=
	\pi\ell_1 r_h\,\widehat{\delta}r_h .
\end{equation}
Integrating on this restricted slice gives
\begin{equation}
	\DeltaS
	=
	\frac{\ell_1}{2}\pi r_h^2,
\end{equation}
where the integration constant is chosen to vanish in the Einstein limit. Combining these contributions gives the entropy entering the first law,
\begin{equation}
	\SH=\SW+\Sone+\DeltaS
	=
	(1+\ell_1)\pi r_h^2 .
\end{equation}
The asymptotic surface charge gives the energy
\begin{equation}
	E=\sqrt{1+\ell_1}\,M,
\end{equation}
and the restricted thermodynamic first law is then satisfied as
\begin{equation}
	\widehat{\delta}E
	=
	T\widehat{\delta}\SH .
\end{equation}

\smallskip
\paragraph{Lightlike branch.}
The lightlike bumblebee configuration is taken as~\cite{Liu:2025oho}
\begin{equation}
	\label{bt0}
	b_t(r)=c,\qquad
	b_r(r)=\sqrt{\frac{(1+\ell_2)b_t^2(r)}{A(r)^2}} .
\end{equation}
with
\begin{equation}
	A(r)=1-\frac{2M}{r},
	S(r)=\frac{1+\ell_2}{A(r)},
	\ell_2=\xi_B c^2.
\end{equation}
The constant \(c\) fixes the chosen lightlike bumblebee background, so \(\ell_2\) is held fixed and \(\widehat{\delta}\ell_2=0\). The surface gravity is
\begin{equation}
	\kappa
	=
	\left.
	\frac{A'(r)}{2\sqrt{1+\ell_2}}
	\right|_{r=r_h}
	=
	\frac{1}{2r_h\sqrt{1+\ell_2}} .
\end{equation}

The radial component again diverges at the horizon in the static frame, so the regular bifurcation surface assumptions required by the standard Iyer--Wald reduction are not satisfied. The Wald entropy and Noether charge correction are
\begin{equation}
	\SW=\pi r_h^2,
	\qquad
	\Sone=-2\ell_2\pi r_h^2 .
\end{equation}
Unlike the spacelike branch, both restricted remainders are nonzero:
\begin{equation}
	\RhX=\frac{\ell_2 \,\widehat{\delta}r_h}{2\sqrt{1+\ell_2}},
	\qquad
	\RhW=\frac{\ell_2 \,\widehat{\delta}r_h}{2\sqrt{1+\ell_2}},
\end{equation}
and the restricted remainders obey
\begin{equation}
	\frac{2\pi}{\kappa}\left(\RhX+\RhW\right)
	=
	\widehat{\delta}\!\DeltaS
	=
	4\pi\ell_2 r_h\,\widehat{\delta}r_h .
	\label{eq:bumblebee-lightlike-RxRw}
\end{equation}
Integrating the restricted one-form, with no bumblebee background work term included, gives
\begin{equation}
	\DeltaS
	=
	2\ell_2\pi r_h^2 .
\end{equation}
Combining \(\SW\), \(\Sone\), and \(\DeltaS\) therefore gives
\begin{equation}
	\SH
	=
	\SW+\Sone+\DeltaS
	=
	\pi r_h^2 .
\end{equation}
The Noether charge correction and \(\DeltaS\) cancel in the final entropy. The asymptotic surface charge gives
\begin{equation}
	E=\frac{M}{\sqrt{1+\ell_2}},
\end{equation}
and it satisfies the restricted thermodynamic first law
\begin{equation}
	\widehat{\delta}E
	=
	T\widehat{\delta}\SH .
\end{equation}
Thus the spacelike branch has \(\Sone=0\) and nonzero \(\DeltaS\), whereas the lightlike branch has nonzero \(\Sone\) and \(\DeltaS\) that cancel in \(\SH\).

\subsection{Extended Gauss--Bonnet gravity}

Gauss--Bonnet gravity in Weyl geometry has attracted recent interest as a
higher-curvature setting in which vector-tensor structures arise naturally.
In Weyl geometry the connection is torsionless but not metric compatible;
the nonmetricity is encoded in a Weyl vector \(W_\mu\). The extended
Gauss--Bonnet construction uses curvature invariants built from the Weyl
connection and leads, after rewriting in Riemannian variables, to a vector-tensor theory with Gauss--Bonnet interactions \cite{BeltranJimenez:2014iie}.
In \(d\)-dimensional spacetime we take~\cite{BeltranJimenez:2014iie,Bahamonde:2025qtc}
\begin{equation}
	S
	=
	\frac{1}{2\kappa_g}\int d^dx\sqrt{-g}\left(
	R-2\Lambda+\alpha\mathcal{G}+\beta\mathcal{G}_{w}
	\right),
	\label{eq:egb-action}
\end{equation}
where
\begin{equation}
	\mathcal{G}
	=
	R_{\mu\nu\rho\sigma}R^{\mu\nu\rho\sigma}
	-4R_{\mu\nu}R^{\mu\nu}
	+R^2
\end{equation}
is the usual Gauss--Bonnet invariant \cite{Lovelock:1971yv}. The Weyl sector correction is
\begin{equation}
	\label{eq:egb-Gw}
	\begin{split}
		\mathcal{G}_{w}
		&=(d-4)(d-3)\Big[
		4G^{\mu\nu}W_{\mu}W_{\nu}
		\\
		&\qquad
		+(d-2)\left(
		4W^{2}\nabla_{\mu}W^{\mu}
		+(d-1)W^{4}
		\right)
		\Big],
	\end{split}
\end{equation}
with \(W^2=W_\mu W^\mu\). Exact black hole solutions of this theory were obtained in Refs.~\cite{Bahamonde:2025qtc,Charmousis:2025jpx,Liu:2025dqg,Alkac:2025zzi}. These solutions 
provide the higher-curvature examples with a Weyl vector used below.

For the horizon calculation, the directly varied presymplectic potential is
\begin{equation}
	\label{eq:egb-theta}
	\begin{split}
		&\Thetaform[\Phi,\delta\Phi]_{bcd}
		=
		\Big(
		2{E_R}^{aefh}\nabla_{h}\delta g_{ef}
		-2\left(\nabla_{h}{E_R}^{aefh}\right)\delta g_{ef}
		\\
		&+\frac{2\beta(d-4)(d-3)(d-2)}{\kappa_g} W^2 g^{ae}\delta W_{e}
		\Big)\varepsilon_{abcd},
	\end{split}
\end{equation}
where
\begin{align}
	{E_R}^{abcd}
	&=
	\frac{1}{2\kappa_g}\Big[
	\left(
	1-\frac{(d-3)(d-4)\beta}{2}W^2
	\right)X^{abcd}
	\notag\\
	&\qquad
	+\beta(d-3)(d-4)W^eW^f{Y_{ef}}^{abcd}
	\notag\\
	&\qquad
	+2\alpha\left(
	R X^{abcd}
	-4R^{ef}{Y_{ef}}^{abcd}
	+R^{abcd}
	\right)
	\Big],
	\label{eq:egb-ER}
	\\
	X^{abcd}
	&=
	g^{a[c}g^{d]b},
	\\
	{Y_{ef}}^{abcd}
	&=
	\frac{1}{2}\bigl(
	{g_{(e}}^a{g_{f)}}^{[c}g^{d]b}- {g_{(e}}^b{g_{f)}}^{[c}g^{d]a}
	\bigr).
\end{align}
The corresponding Noether charge is
\begin{align}
	(\Qxi)_{cd}
	&=
	\Big(
	-{E_R}^{abef}\nabla_{e}\xi_{f}-2\xi_{e}\nabla_{f}{E_R}^{abef}
	\notag\\
	&\quad
	-\frac{2\beta(d-4)(d-3)(d-2)W^2}{\kappa_g}g^{ab}W_{f}\xi^{f}\Big)\varepsilon_{abcd}.
	\label{eq:egb-Qxi}
\end{align}

For the black hole solutions used below we set \(d=5\). The static ansatz is
\begin{align}
	ds^2
	&=
	-A(r)\,dt^2+S(r)\,dr^2+r^2d\Omega_3^2,
	\notag\\
	W_\mu dx^\mu
	&=
	w_0(r)\,dt+w_1(r)\,dr.
\end{align}
For the black holes considered below, \(w_1(r)=w_0(r)/A(r)\). The radial Weyl
component can therefore lead to the same horizon-regularity obstruction as in
the vector-field examples above. Although the solution parameter \(\eta\) may
be promoted to a thermodynamic variable, the horizon integrations are first
carried out on the restricted slice \(\widehat{\delta}\eta=0\).

\paragraph{Case A: \texorpdfstring{\(\alpha=\beta\)}{alpha=beta}.}

When the two couplings coincide and \(\Lambda=0\), the five-dimensional black
hole solution can be written as~\cite{Bahamonde:2025qtc}
\begin{align}
	A(r)
	&=
	1+
	\frac{
		-\frac{m}{r^2}
		-\frac{\alpha\eta^2}{2r^6}
	}{
		1-\frac{2\alpha\eta}{r^4}
	},
	\notag\\
	S(r)
	&=
	\frac{1}{A(r)},
	\notag\\
	w_0(r)
	&=
	\frac{
		-2r^2A(r)+2r^2+\eta
	}{
		4r^3
	},
	\notag\\
	w_1(r)
	&=
	\frac{w_0(r)}{A(r)}.
\end{align}

The horizon radius \(r_h\) is defined as the largest positive root of
\(A(r_h)=0\). We use this condition to trade the integration constant \(m\)
for \(r_h\), obtaining
\begin{equation}
	m
	=
	r_h^2
	-\frac{2\alpha\eta}{r_h^2}
	-\frac{\alpha\eta^2}{2r_h^4}.
	\label{eq:case-A-m-rh}
\end{equation}
Since \(S(r)=1/A(r)\), the surface gravity is
\begin{equation}
	\kappa
	=
	\frac{
		r_h^6+2\alpha\eta r_h^2+\alpha\eta^2
	}{
		r_h^7-2\alpha\eta r_h^3
	}.
\end{equation}

The Wald entropy reads as
\begin{equation}
	\SW
	=
	\frac{\pi^2}{2}\,r_h\left(r_h^2+12\alpha\right),
\end{equation}
and the Noether charge correction is
\begin{equation}
	\Sone
	=
	-\frac{
		3\pi^2\alpha
		\left(2r_h^2+\eta\right)^2
		\left(r_h^4-2\alpha\eta\right)
	}{
		8(	r_h^7+2\alpha\eta\,r_h^3+\alpha\eta^2 r_h )
	}.
	\label{eq:egb5-S1}
\end{equation}

In this case neither restricted horizon remainder vanishes. Direct evaluation of the two pieces gives the one-forms
\begin{equation}
	\RhX
	=
	\frac{
		3\pi\alpha(2r_h^2+\eta)^2(r_h^4+2\alpha\eta)
	}{
		16r_h^5(r_h^4-2\alpha\eta)
	}\,
	\widehat{\delta}r_h,
\end{equation}
and
\begin{equation}
	\begin{split}
		\RhW
		&=
		-\frac{
			3\pi\alpha(2r_h^2+\eta)
		}{
			8r_h^5(r_h^4-2\alpha\eta)
			(r_h^6+2\alpha\eta r_h^2+\alpha\eta^2)
		}
		\\
		&\quad\times
		\Big[
		4r_h^{12}
		+2r_h^{10}\eta
		+4\alpha\eta r_h^8
		-4\alpha\eta^2 r_h^6
		\\
		&\qquad
		+\alpha\eta^2(24\alpha-\eta)r_h^4
		+20\alpha^2\eta^3 r_h^2
		+4\alpha^2\eta^4
		\Big]\widehat{\delta}r_h .
	\end{split}
\end{equation}
With the normalization used in Eq.~\eqref{eq:xi-theta-decomp-main}, the restricted sum is constrained by the entropy identity as
\begin{equation}
	\frac{2\pi}{\kappa}\left(\RhX+\RhW\right)
	=
	\widehat{\delta}\!\DeltaS .
	\label{eq:egb5-RxRw}
\end{equation}
The restricted integration with \(\widehat{\delta}\eta=0\) fixes \(\Delta S\)
up to an additive function \(F(\eta)\). We use the natural normalization
\(F(\eta)=0\), since a nonzero \(F\) would only be absorbed into a redefinition
of the conjugate potential \(\Phi_\eta\). With this normalization,
\begin{equation}
	\label{eq:egb5-DeltaS}
	\begin{split}
		\DeltaS
		&=
		\frac{3\pi^2\alpha}{8(	r_h^7+2\alpha\eta\,r_h^3+\alpha\eta^2 r_h )}
		\Big[
		-12r_h^8
		+12r_h^6\eta
		\\
		&\qquad
		-r_h^4\eta(40\alpha-\eta)
		-8r_h^2\alpha\eta^2
		+6\alpha\eta^3
		\Big].
	\end{split}
\end{equation}
\begin{table*}[t]
	\centering
	\def\leftlegendglue{\hfil}
	\caption{\label{tab:summary-models}Summary of the illustrative models and their entropy corrections.}
	\scriptsize
	\setlength{\tabcolsep}{3pt}
	\begin{tabular}{@{}p{0.30\textwidth}cccp{0.08\textwidth}p{0.24\textwidth}@{}}
		\toprule
		Model or branch & \(\Sone\) & \(\DeltaS\) & \(\RhX\) & \(\RhW\) & Outcome \\
		\midrule
		Kalb--Ramond & \(0\) & \(0\) & \(0\) & \(0\) & \(\SH=\SW\) \\
		Bumblebee (spacelike VEV) & \(0\) & \(\neq0\) & \(0\) & \(\neq0\) &  \(\SH\neq\SW\) \\
		Bumblebee (lightlike VEV) & \(\neq0\) & \(\neq0\) & \(\neq0\) & \(\neq0\) & \(\SH=\SW\)   (\(\Sone+\DeltaS=0\) )\\
		Extended Gauss--Bonnet, Case A (\(\alpha=\beta\)) & \(\neq0\) & \(\neq0\) & \(\neq0\) & \(\neq0\) & \(\SH\neq\SW\) \\
		Extended Gauss--Bonnet, Case B (\(\alpha\neq\beta\)) & \(\neq0\) & \(\neq0\) & \(\neq0\) & \(\neq0\) & \(\SH\neq\SW\) \\
		\bottomrule
	\end{tabular}
\end{table*}
With the \(S_1\) in Eq.~\eqref{eq:egb5-S1}, this gives
\begin{equation}
	\label{eq:egb5-total-entropy}
	\begin{split}
		\SH
		&=
		\SW+\Sone+\DeltaS
		\\
		&=
		\frac{\pi^2}{2}
		\left(
		r_h^3+\frac{6\alpha\eta}{r_h}
		\right).
	\end{split}
\end{equation}
The thermodynamic interpretation is then checked in the enlarged phase space. The energy is obtained from the integrable asymptotic surface charge \(\delta H_\xi^{(\infty)}\). For this five-dimensional branch,
\begin{equation}
	E=\frac{3\pi}{8}m .
\end{equation}
Treating \(\eta\) as an independent solution parameter, the conjugate potential is
\begin{equation}
	\label{eq:egb5-Phi-eta}
	\Phi_\eta
	=
	-\frac{3\pi\alpha}{4}
	\left(
	\frac{1}{r_h^2}
	+\frac{\eta}{2r_h^4}
	+\frac{2 \kappa}{r_h}
	\right).
\end{equation}
The enlarged thermodynamic first law takes the form
\begin{equation}
	\label{eq:egb5-first-law-eta}
	\delta E
	=
	T\,\delta \SH
	+\Phi_\eta\,\delta\eta .
\end{equation}
\paragraph{Case B: \texorpdfstring{\(\alpha\neq\beta\)}{alpha not equal beta}.}

For unequal couplings, we set \(\Lambda=0\). The five-dimensional solution has
two algebraic branches labelled by \(\sigma=\pm1\). We use the \(\sigma=-1\)
branch~\cite{Bahamonde:2025qtc},
\begin{align}
	A(r)
	&=
	1-\frac{\beta\eta}{2(\alpha-\beta)r^2}
	\notag\\
	&\quad
	+\frac{r^{2}}{4(\alpha-\beta)}
	\Bigg[
	1
	\notag\\
	&\qquad
	-\sqrt{
		1
		+\frac{2\bigl(4(\alpha-\beta)m-2\beta\eta\bigr)}{r^4}
		+\frac{4\alpha\beta\eta^2}{r^8}
	}
	\Bigg],
	\notag\\
	S(r)
	&=
	\frac{1}{A(r)},
	\notag\\
	w_0(r)
	&=
	\frac{
		-2r^2A(r)+2r^2+\eta
	}{
		4r^3
	},
	\notag\\
	w_1(r)
	&=
	\frac{w_0(r)}{A(r)}.
	\label{eq:egb-case-B}
\end{align}
The horizon radius \(r_h\) is defined as the largest positive root of
\(A(r_h)=0\). In what follows we trade the integration constant \(m\)
for \(r_h\):
\begin{equation}
	m
	=
	r_h^2+2(\alpha-\beta)
	-\frac{2\beta\eta}{r_h^2}
	-\frac{\beta\eta^2}{2r_h^4}.
	\label{eq:case-B-m-rh}
\end{equation}
All horizon quantities below are understood after this substitution.
Since \(S(r)=1/A(r)\), the surface gravity is
\begin{equation}
	\kappa
	=
	\frac{
		r_h^6+2\beta\eta r_h^2+\beta\eta^2
	}{
		r_h^3\left[
		r_h^4+4r_h^2(\alpha-\beta)-2\beta\eta
		\right]
	}.
\end{equation}
The Wald entropy is independent of \(\beta\) at this stage,
\begin{equation}
	\label{eq:egb5B-SW}
	\SW
	=
	\frac{\pi^2}{2}\,
	r_h\left(r_h^2+12\alpha\right).
\end{equation}

The non-Wald Noether charge correction is
\begin{equation}
	\label{eq:egb5B-S1}
	\Sone
	=
	-\frac{
		3\pi^2\beta
		\left(2r_h^2+\eta\right)^2
		\left[
		r_h^4
		+4r_h^2(\alpha-\beta)
		-2\beta\eta
		\right]
	}{
		8 (r_h^7+2\beta\eta r_h^3+\beta\eta^2 r_h )
	}.
\end{equation}
As in Case A, neither restricted horizon remainder vanishes on this branch. The restricted entropy identity gives
\begin{equation}
	\frac{2\pi}{\kappa}\left(\RhX+\RhW\right)
	=
	\widehat{\delta}\!\DeltaS ,
	\label{eq:egb5B-RxRw}
\end{equation}
with \(\widehat{\delta}\eta=0\) and fixed couplings. The restricted integration again determines \(\DeltaS\) up to an additive function of \(\eta\), which we set to zero with the same normalization convention. Integrating it gives
\begin{equation}
	\label{eq:egb5B-DeltaS}
	\begin{split}
		\DeltaS
		&=
		\frac{3\pi^2\beta}{8(r_h^7+2\beta\eta r_h^3+\beta\eta^2 r_h )}
		\Big[
		-12r_h^8\\
		&\qquad+4r_h^6(4\alpha-4\beta+3\eta)
		+r_h^4\eta(16\alpha-56\beta+\eta)
		\\
		&\qquad
		+4r_h^2(\alpha-3\beta)\eta^2
		+6\beta\eta^3
		\Big].
	\end{split}
\end{equation}
Combining the three entropy contributions gives
\begin{equation}
	\label{eq:egb5B-total-entropy}
	\begin{split}
		\SH
		&=
		\SW+\Sone+\DeltaS
		\\
		&=
		\frac{\pi^2}{2}
		\left(
		{r_h^3}
		+12r_h(\alpha-\beta)
		+\frac{6\beta\eta}{r_h}
		\right).
	\end{split}
\end{equation}
The thermodynamic interpretation is again checked only after \(\SH\) has been assembled. The energy follows from the same asymptotic surface charge,
\begin{equation}
	E=\frac{3\pi}{8}m .
\end{equation}
The conjugate potential for the five-dimensional unequal-coupling branch is
\begin{equation}
	\label{eq:egb5B-Phi-eta}
	\Phi_\eta
	=
	-\frac{3\pi\beta}{4}
	\left(
	\frac{1}{r_h^2}
	+\frac{\eta}{2r_h^4}
	+\frac{2 \kappa}{r_h}
	\right),
\end{equation}
The enlarged thermodynamic first law then reads
\begin{equation}
	\label{eq:egb5B-first-law-eta}
	\delta E
	=
	T\,\delta \SH
	+\Phi_\eta\,\delta\eta .
\end{equation}

These two branches show how the Weyl vector and Gauss--Bonnet terms enter \(E_R^{abcd}\), \(\ThetaS\), and the entropy-sector Noether charge \(\QS\). They provide the higher-curvature counterpart of the bumblebee example, with both the non-Wald charge term and the remaining horizon contribution active.

The outcomes of the illustrative models are summarized in Table~\ref{tab:summary-models}.

\section{Conclusion}
\label{sec:conclusion}
For the stationary black hole solutions studied here, the entropy entering the first law decomposes as
\begin{equation}
	\SH=\SW+\Sone+\DeltaS .
\end{equation}
The two terms beyond the Wald density are the entropy-sector Noether charge correction \(\Sone\) and the remaining contribution \(\DeltaS\) from the entropy-sector horizon surface charge variation.

At the level of the horizon integrand, the key decomposition is
\begin{equation}
	-\oint_{\Bhor} \xi\cdot\ThetaS
	=
	-\frac{\SQ}{2\pi}\,\delta\kappa+\mathcal R_X+\mathcal R_W .
\end{equation}
The term \(\mathcal R_X\) comes from those horizon components of
\(E_R^{abcd}\nabla_d\delta g_{bc}\) that are not contained in the Wald entropy density.
The term \(\mathcal R_W\) is the \(W\)-part remainder after removing the
\(-\Sone\delta\kappa/(2\pi)\) part fixed by \(\mathbf W_a\xi^a\).

The examples realize different combinations of these two sources. The Kalb--Ramond solution has no additional contribution and \(\SH=\SW\). In bumblebee gravity, the spacelike branch has \(\Sone=0\) but \(\DeltaS\neq0\), while in the lightlike branch both \(\Sone\) and \(\DeltaS\) are nonzero and cancel in \(\SH\). In the five-dimensional extended Gauss--Bonnet branches studied here, both sources are active and the enlarged thermodynamics is \(\delta E=T\delta \SH+\Phi_\eta\delta\eta\), with \(E=(3\pi/8)m\) from \(\delta H_\xi^{(\infty)}\).

The result refines the Iyer--Wald analysis for solution families where the usual bifurcation surface argument does not by itself settle the entropy. When the standard regularity assumptions hold, the additional contributions vanish. When they fail, the criterion identifies whether the correction is already contained in the entropy-sector Noether charge, in the remaining \(\xi\cdot\ThetaS\) term, or in both. A JKM transformation can move terms between \(\Qxi\) and \(\Thetaform\), and therefore can change how the correction is divided between \(\Sone\) and \(\DeltaS\).

This criterion separates regular tensor backgrounds, singular vector backgrounds, and higher-curvature examples with a Weyl vector in a common covariant phase space language. It would be useful to extend the analysis to rotating families, genuinely
nonstationary horizons, and more general enlarged phase spaces. For nonminimally coupled black holes of the type considered here, this analysis can also be combined with the extended Iyer--Wald formalism and the universal thermodynamic-volume construction of Refs.~\cite{Xiao:2023lap,Xiao:2025gle}.
\section*{Acknowledgments}
This work was supported by the National Natural Science Foundation of China (Grants No. 12475056, No. 12475055, No. 12247101), the 111 Project (Grant No. B20063), the Fundamental Research
Funds for the Central Universities 
(Grants No. lzujbky-2025-jdzx07), 
 the Natural Science
Foundation of Gansu Province (No. 
22JR5RA389 and
No. 25JRRA799), and Gansu Province's Top Leading Talent Support Plan.
\bibliography{wald_entropy_mod_en_refs}

\end{document}